\documentclass[useAMS,usegraphicx,usenatbib]{mn2e}

\usepackage{amssymb}
\usepackage{amsmath}
\usepackage{color}

\newcommand{\be}{\begin{equation}}
\newcommand{\ee}{\end{equation}}
\newcommand{\kb}{$k_{b}$\,}
\newcommand{\mpr}{$m_{p}$\,}

\newcommand{\mvir}{{\rm M}$_{v}$\,}

\newcommand{\rnfw}{{\rm r}$_{v}$\,}
\newcommand{\mbh}{{\rm M}$_{bh}$\,}
\newcommand{\rhoism}{$\rho_{c0}$\,}

\newcommand{\vphi}{{\rm v}$_{\phi}$\,}

\title[AGN feedback]{AGN activity: self-regulation from backflow}
\author[Antonuccio-Delogu \& Silk]{\noindent 
V. Antonuccio-Delogu$^{1,2,3}$, Joseph Silk$^{1}$\\
$^{1}$Astrophysics, Department of Physics, University of Oxford, Keble Road Ox1 3RH, Oxford, United Kingdom\\
$^{2}$INAF - Osservatorio Astrofisico di Catania, Via S. Sofia 78, Catania, I-95123, Italy\\
$^{3}$Scuola Superiore di Catania, Via San Nullo 5/i, Catania, I-95123, Italy }

\begin{document} 

\date{Accepted ??. Received ??; in original form 2009 ??}

\pagerange{\pageref{firstpage}--\pageref{lastpage}} \pubyear{2007}

\maketitle
%\draftcopySetGrey 0.6

\label{firstpage}

\begin{abstract}
  We study the internal circulation within the cocoon carved out by the relativistic jet emanating from an AGN within the ISM of its host galaxy. Firstly, we develop a model for the origin of the internal flow, noticing that a significant increase of large scale velocity \emph{circulation} within the cocoon arises as significant gradients in the density and entropy are created near the hot spot (a consequence of Crocco's vorticity  generation theorem). We find simple and accurate approximate solutions for the large scale flow,showing that a backflow towards the few inner parsec region develops. We solve the appropriate fluid dynamic equations, and we use these solutions to predict the mass inflow rates towards the central regions.\\
\noindent 
We then perform a series of 2D simulations of the propagation of jets using FLASH 2.5, in order to validate the predictions of our model. In these simulations, we vary the mechanical input power between $10^{43}$ and $10^{45}$
  ergs$\cdot {\rm sec}^{-1}$, and assume a NFW
  density profile for the dark matter halo, within which a isothermal
  diffuse ISM is embedded. The backflows which arise supply the central AGN region with very low angular momentum gas, at average rates of the order of $0.1-0.8\, \rm{M}_{\odot}\, \rm{yr.}^{-1}$, the exact value seen to be strongly dependent on the central ISM density (for fixed input jet power). The time scales of these inflows are apparently weakly dependent on the jet/ISM parameters, and are of the order of $3-5\times 10^{7}\, \rm{yrs}$.\\
We then argue that these backflows could (at least partially) feed the AGN, and provide a self-regulatory mechanism of AGN activity, that is not directly controlled by, but instead controls,  the star formation rate within the central circumnuclear disk.
 
\end{abstract}


\section{Introduction}
There is significant evidence for a close connection between the presence of compact objects (hereafter \emph{Black Holes}, BHs) in the dense central regions of most galaxies and the AGN phenomenon. Moreover, since the original theoretical suggestion by \citet{1998A&A...331L...1S}, significant observational evidence  accumulated concerning the impact of nuclear activity on the global stellar evolution within the host galaxy \citep[see e.g.][for some recent work]{2006Natur.442..888S, 2007RMxAC..28..109Y, 2007MNRAS.382..960K, 2008MNRAS.388...67K}. In addition to its effect on \emph{global} star formation, the presence of an AGN seems  to also be connected with \emph{circumnuclear starbursts} on small (from -parsec to kilo-parsec) scales \citep{2001ApJ...559..147S, 2005ASSL..329..263G, 2007MNRAS.380..949S, 2007ApJ...671.1388D}. However, despite all this observational evidence, the connection between AGN activity, negative/positive feedback on star formation, and local starbursts is not well understood.\\
\noindent
If the main physical agent for this connection is the interaction between the jet and the host galaxy's interstellar medium (hereafter ISM), then the modeling of the propagation of AGN relativistic jets into the ISM should have as a primary target the understanding of the jet interaction with existing stellar populations, and the mechanisms which feed the parsec-scale accretion disc around the central BHs. Recent simulations \citep{2007ApJS..173...37S, 2008MNRAS.389.1750A} have begun to  self-consistently model  the feedback of a relativistic jet on its host galaxy, and the global consequences for e.g. blue-to-red cloud migration and downsizing \citep{2009MNRAS.396...61T}. \\
\noindent
In this work, we show that the propagation of the AGN jet into the ISM not only has an impact on large-scale star formation within the host galaxy, but also on the feeding of the parsec-scale accretion disc around the BH  in the very central region of the AGN. We show that the generation of steep density  and entropy gradients near the \emph{recollimation shock} (hereafter RS) and the \emph{hot spot} (hereafter HS) induce a \emph{backflow} within the cocoon, which in turn bends back towards the central region, thus providing a secondary accretion flow which is capable of feeding the central BH at rates of the order of a few $\rm{M}_{\odot}\, \rm{yr.}^{-1}$, over timescales of a few times $10^{7}\, \rm{yrs.}$ Thus, this backflow could easily provide the central BH with the ''fuel'' needed to support the accretion disc and the production of the jet, until the RS is destroyed and the backflow disappears.\\
\noindent
This paper is organised as follows. In section 2 we develop a model for the origin of the backflow, by applying an exact theorem from fluid dynamics which connects the \emph{circulation} to the entropy gradients (\emph{Crocco's theorem}). In section 3 we describe the results of a series of simulations aimed at validating the model developed in section 2, and we also discuss the implications for the feeding of the central BH. We draw our conclusions in section 4.\\
\noindent 
In the following, we express lengths in units of of kiloparsecs,
but  generally,  we adopt cgs units, unless otherwise 
explicitly stated. The underlying
cosmological parameters are taken from the 3-year WMAP best fit
$\Lambda$CDM model \citep{2006NewAR..50..850B}: ${\rm H}_{0} =
74\pm 3$ Km/sec/Mpc, $\Omega_{m} = 0.234\pm 0.035$,
$\Omega_{b}h^{2} = 0.0223\pm 0.0008$. The unit of time is taken to be
the Hubble time for this cosmology, i.e.: $t_{0} = 1.365\times 10^{10}$ years. $G$, \kb and \mpr denote the
gravitational constant, Boltzmann constant and proton mass, respectively.
\section[]{Formation of the backflow} \label{sect:2}

\subsection{Model}
\begin{figure*}
%\centering
\includegraphics[scale=0.5,angle=0]{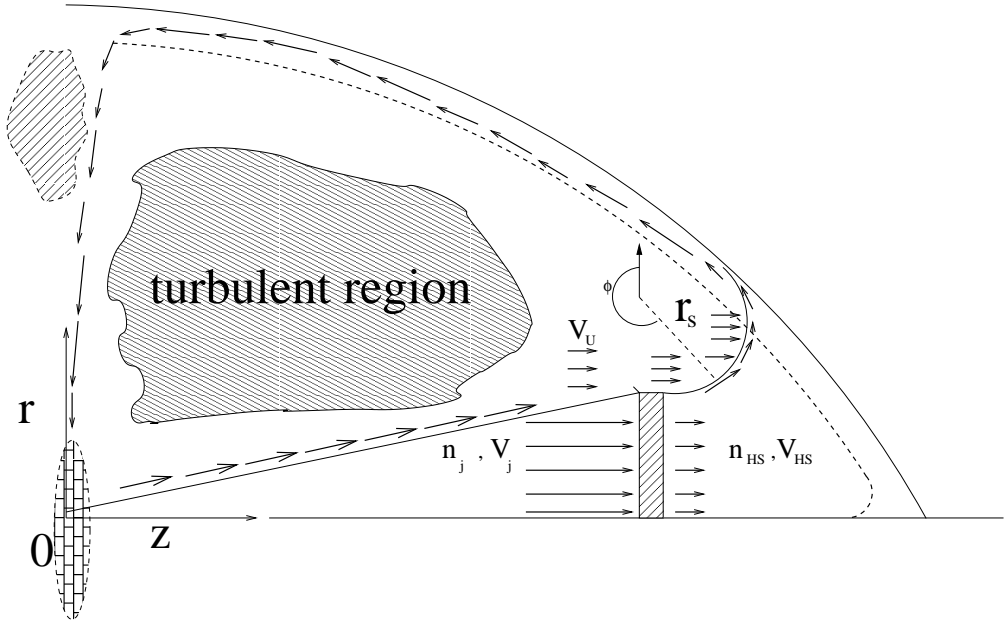}
\caption{Schematic model of the origin of circulation near the hot spot
and in the equatorial plane of the cocoon. We assume that the
\emph{hotspot} (HS) is bounded by a lateral spherical section having
curvature radius $r_{s}$.}
%\label{fig_fr1}
\label{figm1}
\end{figure*}

 The model we propose here for the origin and evolution of the circulation within the
 cocoon extends the models by
 \citet{1991MNRAS.250..581F} and \citet{1997MNRAS.286..215K}. These
 papers show that a \emph{recollimation shock} (hereafter RS) forms at
 some distance along the path of the jet, when the latter is confined by the cocoon's pressure. The post-shocked gas then accumulates into a terminal region confined between the RS and the outer surface of the bow shock(see Figure~\ref{figm1}), with a density $n_{hs}
 \approx 7 n_{j}$, as appropriate to shocked gas arising from a relativistic equation of state. We generically call this region the \emph{hot spot} (HS).
\noindent

In the original model by Fall, the only systematic flow which is present is that within the jet: the cocoon is assumed to be filled with turbulent gas, and the only systematic motion within it is assumed to be its expansion within the host galaxy's ISM. Moreover, the jet is assumed to be completely pressure-confined by the cocoon. Under these conditions, \citet{1991MNRAS.250..581F} and \citet{1997MNRAS.286..215K} show that a recollimation shock near the injection point of the jet develops. \\
\noindent
However, during the initial phases, when the pressure within the cocoon is not yet sufficient to confine the jet, the latter can expand into the cocoon and acquires a finite opening angle (see Figure~\ref{figm1}). We study in detail this phase in this paper. If there is an initial transverse velocity gradient ($\partial v_{z}/\partial r \neq 0$, the central parts of the jet are supersonic and form a shock, while the flow in the outermost part is subsonic. The shocked gas will accumulate in a terminal region called the \emph{Hot Spot} (hereafter HS). The lateral, subsonic flow will eventually also reach the HS and, as it crosses it, will acquire an angular momentum and will be deviated back, flowing along the inner boundary of the bow shock.\\
\noindent
During a later phase, when the pressure in the cocoon has grown large enough, the jet is pressure-confined and the usual scenario applies: the recollimation shock now is nearer to the injection point, as originally predicted by \citet{1997MNRAS.286..215K}. The lateral flow now reaches the HS and is simply reflectd back, generating a backflow which does not propagate along the bow shock, but is flowing along the z-axis, towards the origin. This has also been seen in a previous work \citep{1999MNRAS.305..707K}, including the simulations we present here (Figures~\ref{fig1b} and ~\ref{fig2}).\\
\noindent 
During both phases, the cocoon is occupied by a very low density, high temperature gas. In this region the gas is in a turbulent state, with very
little or no systematic motions.\\
\noindent
Obviously, the shocked gas within the hot spot has a higher entropy
than that in the cocoon. Under these conditions, \emph{Crocco's theorem} states that  a circulation arises  within a compressible fluid, even if in laminar motion
 \citep[see][for a more recent discussion]{2001...2100..25stzmatost}. In a planar geometry one obtains:
\be
\omega v = T\frac{ds}{dn} - \frac{dh_{0}}{dn}
\label{eq:bckflw:1}
\ee
where: $n$ is the normal to the direction perpendicular to the streamline, $\omega\equiv\mid\boldsymbol{\omega}\mid$ is the modulus of the circulation, and $s$ and $h_{0} = h +
v^{2}/2$ are the specific entropy and stagnation enthalpy,
respectively.\\
\noindent
\subsubsection{Flow near the Hot Spot}
\begin{figure}
\centering
\includegraphics[scale=0.45,angle=0]{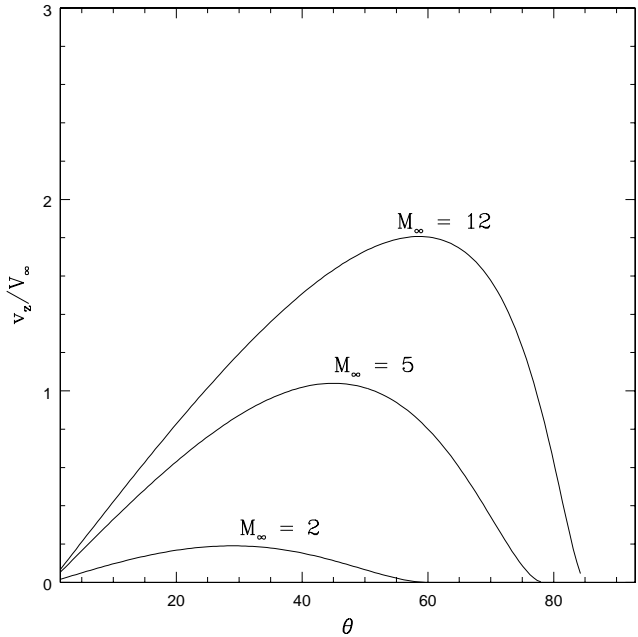}
\caption{The excursion of $\mid v_{d\mid z}/v_{u\mid z}\mid$, as
  measured in eq.~\ref{eq:bckflw:3}, for different values of the
  upstream Mach number. The flows within the cocoon usually are
  transonic, with $M_{\infty} \simeq 1-2$.}
%\label{fig_fr1}
\label{fig_v_theta}
\end{figure}

We will first consider the change in circulation induced by the presence
of the reconfinement shock. We will approximate the recollimation shock surface as planar, thus the entropy is constant across the streamline, and the flow will not gain any circulation. However, the gas flowing
\emph{near and outside} the jet will also eventually reach the HS: the boundary layer separating this region from the turbulent region will be a curved surface, joining the recollimation shock to the bow shock (see Fig.~\ref{figm1}). We then assume that this off-axis flow motion is predominantly directed along the $z-$
direction, thus: $\mathbf{v} \equiv v_{u\mid z}\hat{\mathbf{z}}$. Hereafter, subscripts $u$ and $d$ will denote
quantities computed in the upstream and downstream regions,
respectively, where by the latter we mean the HS, as shown in
Figure~\ref{figm1}. In order to compute $\omega_{hs}$, the
circulation in the hot spot, we will make use of the expression for
the circulation near a curved shock, derived by \citet[][eq. 8]{1958ZAMP...9..637}:
\be
\omega_{hs} = \frac{2v_{z\mid u}}{r_{s}\left(\gamma + 
  1\right)}\frac{\left(M_{n}^{2}-1\right)\mid\cos\phi\mid}{M_{n}^{2}[2+\left(\gamma-1\right)M_{n}^{2}]}
\label{eq:bckflw:2}
\ee
 where: $M_{n} = (v_{z}/c_{s})\cos\theta$ is the Mach number in the
 upstream flow region. As we show in Appendix~\ref{append_2a}, one can obtain an exact solution in 2D spherical coordinates for the velocity field after the shocked layer:
\be
v_{\phi}(r_{s}) = \frac{4v_{z\mid u}}{\left(\gamma + 
  1\right)}\frac{\left(M_{n}^{2}-1\right)\mid\cos\phi\mid}{M_{n}^{2}[2+\left(\gamma-1\right)M_{n}^{2}]}
\label{eq:bckflw:3}
\ee 
where $\phi$ is a polar angle (see Fig~\ref{figm1}) and, in the coordinates system shown in Fig.~\ref{figm1}, we have: $v_{z}=-v_{\phi}\cos\phi, v_{r}=v_{\phi}\sin\phi$. We see that the longitudinal z-component of velocity after the shock acquires a negative component, and the polar component in the downstream
region acquires a positive value: thus, a \emph{backflow} develops.\\
\noindent
Note that the magnitude of the velocity variation only depends on $\gamma$ and
$M_{n}$. As we can see from Figure~\ref{fig_v_theta}, this variation is not very large, for values of $M_{n}\approx 1-2$, typical of the transonic flows which develops within the cocoons \citet[][hereafter paper I]{2008MNRAS.389.1750A}.\\
\noindent
\subsubsection{The flow along the bow shock}
\begin{figure}
\centering
\includegraphics[scale=0.45,angle=0]{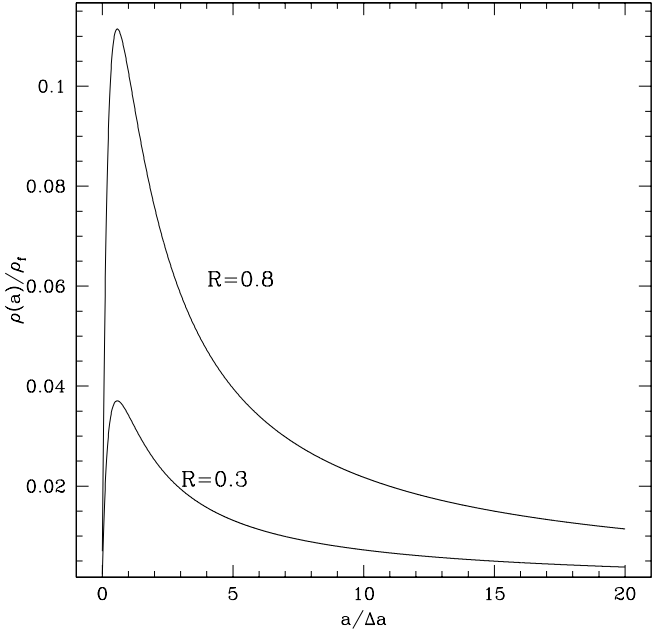}
\caption{The scaled density within the bow shock as a function of
  $a/\Delta a$, the semi-major axis measured in units of the width of
  the spheroidal shell containing the shock. We plot the ratio
  $\rho/(3\rho_{0}a_{0}^{2}/(\Delta a)^{2})$, for two different values
  of the aspect ratio $R$.}
%\label{fig_fr1}
\label{fig_rho_l}
\end{figure}

\begin{figure}
\centering
\includegraphics[scale=0.45,angle=0]{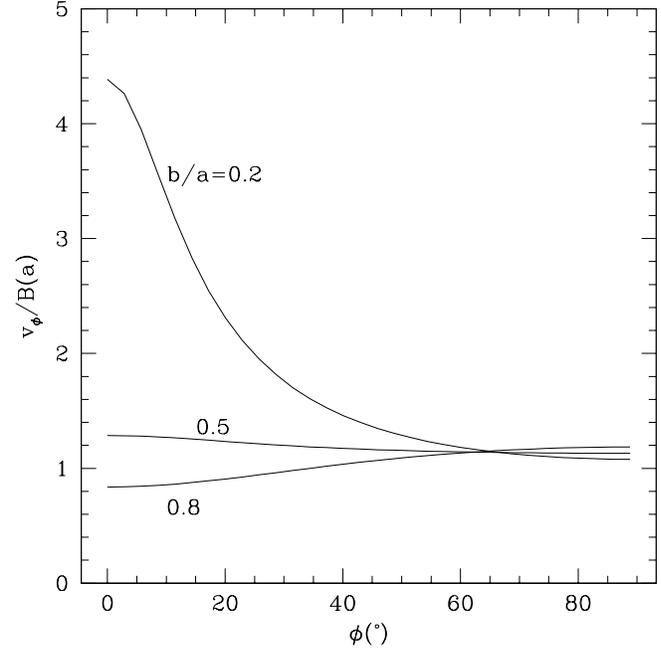}
\caption{The angular factor in eq.~\ref{eq:bckflw:13}, for different
  aspect ratios of the bow shock.}
%\label{fig_fr1}
\label{fig_vphi_fact}
\end{figure}

After the HS, the backflow enters into the inner part of the bow shock (Figure~\ref{figm1}), and the magnitude of
the rotation is determined by the conservation law for vorticity in
2D (see Appendix~\ref{append_1} for a proof): 
\be
\frac{\mid\boldsymbol{\omega}\mid}{\rho} = const
\label{eq:bckflw:4}
\ee
Thus, in order to determine $\mid\boldsymbol{\omega}\mid$, we have to determine the density inside the bow shock. We notice that this
layer can not be uniform, because the density, perpendicular to the jet, must necessarily be higher than
that in the intermediate regions. In fact, for symmetry
reasons, the velocity $v_{z} \approx 0$ near the meridional plane of the bow shock, thus by conservation of the momentum flux, we immediately
see that the density must increase.\\
\noindent 
We will obtain an estimate of the density within this layer by making
three assumptions: a) the bounding surface and its inner part are
approximated as oblate spheroids; b) all the mass swept out since the
formation of the cocoon ends in the bow shock; c) the bow shock
layer has a constant major axis width $\Delta a$. Then the volume of
this spheroidal bow shock layer is given by:
\be
\Delta V = \frac{4\pi}{3}R^{2}\Delta a\left[ 3a_{i}^{3} +
  3a_{i}\delta a + \Delta a^{2}\right]
\label{eq:bckflw:5}
\ee
 where $R=b_{i}/a_{i}$ and $a_{i}$ are the aspect ratio and the length of the inner semi-major axis, respectively. This 
 gives
\be
M(a_{i}) = 4\pi\rho_{0}a_{0}^{2}a_{i}\frac{R}{\sqrt{1-R^{2}}}\arctan(\sqrt{1-R^{2}})
\label{eq:bckflw:7}
\ee
Then, the average density within the spheroidal bow shock layer of thickness
$\Delta a$ can be obtained by dividing $M(a_{i})$ by the volume $\Delta V$:
\be
\rho_{bs} = 3\rho_{0}\left(\frac{a_{0}}{\Delta
  a}\right)^{2}g(R)\frac{\lambda}{3\lambda^{2} + 3\lambda + 1}
\label{eq:bckflw:8}
\ee
where we have defined $\lambda = a_{i}/\Delta a$, and we have dropped the
subscript $i$ from the semi-major axis. Note that the time-dependance
of the density enters through the semi-major axis ($a\equiv a(t)$).\\
\noindent
We notice that the evolution of the density in the bow shock depends only on
the factor $\lambda$: in Figure~\ref{fig_rho_l} the
average density has a peak at around $a\approx 0.85\Delta a$, and then
decreases relatively slowly.\\
\noindent
Knowing the density, we can compute the vorticity of the flow after it
has been deflected by the HS. We will make use of oblate spheroidal coordinates in 2D, which define an orthogonal curvilinear coordinate system \citep{1972hmfw.book.....A}:
\begin{eqnarray}
x & = & a\cosh\xi\cos\psi \\
y & = & a\sinh\xi\sin\psi
\label{app0:eq1}
\end{eqnarray}
As is well known, the scale factors for the $(\xi,\psi)$ coordinates
 are identical: $h_{\xi} = h_{\psi} = a\left(\sinh^{2}\xi +
 \sin^{2}\psi\right)^{1/2}$. Curves of constant $\xi$ are oblate
 spheroids, with axis ratio $b/a = \tanh\xi$. The eccentricity of
 these spheroids is then given by:  $\epsilon = \sqrt{1 - (b/a)^{2}} = 1/\cosh\xi$.\\
\noindent
In these coordinates the vorticity $\boldsymbol{\omega}$ is given by:
\be
\boldsymbol{\omega} \equiv \omega_{z} =
\frac{1}{h_{\xi}h_{\psi}}\left[\frac{\partial(h_{\psi}v_{\xi})}{\partial
    \xi} - \frac{\partial(h_{\xi}v_{\psi})}{\partial\psi}\right]
\label{app0:eq2}
\ee
and the stationary continuity equation can be expressed as:
\be
\nabla\cdot(\rho\boldsymbol{v}) =
\frac{1}{h_{\xi}h_{\psi}}\left[\frac{\partial(h_{\psi}\rho v_{\xi})}{\partial
    \xi} + \frac{\partial(h_{a}\rho v_{\psi})}{\partial\psi}\right] = 0
\label{app0:eq3}
\ee
In Appendix~\ref{append_2b} we use these equations and eq.~\ref{eq:bckflw:4} to deduce the tangential velocity $v_{\psi}$:
\be
v_{\psi}^{2} = \frac{B(a)}{\sinh^{2}\xi+\sin^{2}\psi}\left[\alpha + G(\xi) + 2\cosh\xi\cdot\sin^{2}\psi\right]
\label{eq:bckflw:13}
\ee
where: $B(a) = ac_{0}(a)\rho_{0}v_{0}$, $\alpha$ is a factor depending on the initial condition at the \emph{injection point}, chosen in such a way to guarantee that the term inside the parentheses is positive. Finally, we have defined: 
\[
G(\xi)=\frac{1}{6}\cosh(3\xi)-\frac{3}{2}\cosh\xi
\]
In the oblate spheroidal coordinate system $(\xi,\psi)$ adopted to derive this solution, we have: $\psi_{0}\leq\psi\leq \pi/2$, and the upper limit corresponds to the meridional plane on ellipsoidal surfaces $\xi=const$. We show in Figure~\ref{fig_vphi_fact} the angular variation of $v_{\psi}$, for some values of the eccentricity. As we  see, the
maximum variations are always around unity, along the streamlines,
and the velocity tends to decrease or to stay almost constant, as we approach the polar regions.
\noindent
Correspondingly, the density
must increase or stay almost the same, by flow conservation. The
variation of the velocity is not very large, but depends also on the
average density inside the layer. As
this  holds true for all the flow streamlines contained within the
cross section of the bow shock, we deduce that a \emph{high density
  region} will form perpendicularly to the jet, near the end of the
laterally expanding region, for low values of the aspect ratio. As a consequence, we expect a steep increase of the
vorticity for those streamlines approaching the high density region,
and the flow will consequently bend towards the meridional plane.\\
\noindent 
In summary, our model makes a few predictions. First, at least during
the initial phases of the expansion, a circulation arises inside the
cocoon, induced by the presence of a high density and entropy region (the Hot Spot). The backflow which develops tends to follow the bow
shock, and then it bends back near the meridional axis, perpendicular
to the jet. For symmetry reasons, this flow will converge back towards
the jet.\\
\noindent
Although we have derived expressions for the velocity along a typical
streamline (eqs.~\ref{eq:bckflw:3} and ~\ref{eq:bckflw:13}), these
should be taken to provide only  an order of magnitude estimate of
the actual situation. The presence of an extended turbulent region in
the central region of the cocoon will introduce fluctuations, and we
could think of different factors which would make our formulae just a first order approximation. Thus it is worthwhile  to use  \emph{numerical simulations} to verify
whether this model could provide a realistic description of the
backflow, and the spatial and temporal extent of its validity.
 
\section{Simulations}

\subsection{Simulation setup}
The observational evidence shows that there exist 
correlations between the main physical properties of central BHs, AGN
properties and properties of their host galaxy halos. Some of these correlations, like the relation $\rm{M}_{bh}-\sigma$, have an intrinsic scatter that is relatively small when compared to others (e.g. the $\rm{P}_{jet}-\rm{M}_{bh}$ relationship). Our simulation
setup takes these correlations into account: we have designed a set of
9 simulations where the main parameter we change is the velocity
dispersion of the host DM halo ($\sigma_{v}$), and all the other parameters are
determined by using known scaling relations. We determine the total
virial mass of the host DM halo using the relation found by
\citet{2006ApJ...648..826L}, using a sample from the SDSS DR3:
\be
{\rm M}_{v} = 2.57\times 10^{12} \sigma_{200}^{2.99\pm
    0.15}\left[{\rm km/sec}\right] 
\label{eq:1}
\ee
where $\sigma_{200}$ is the DM central velocity dispersion in units of $200\, \rm{km}\,\rm{sec}^{-1}$. 
\mvir is needed to compute the concentration parameter of the central
NFW halo model, as given by the fit of \citet{2001MNRAS.321..559B}:
\be
c_{v} = 9\left(\frac{M_{v}}{M_{\ast}}\right)^{-0.13}, \, M_{\ast}=
1.5\times 10^{13} h^{-1} M_{\sun} 
\label{eq:2}
\ee
We determine the BH mass \mbh using the relation given by
\citet{2000ApJ...539L...9F}:
\be
{\rm M}_{bh} = (1.2\pm 0.2)\times 10^{8} \sigma_{200}^{3.57\pm 0.3} M_{\sun} 
\label{eq:3}
\ee

Finally, we compute the jet's mechanical power ${\rm P}_{jet}$ using the
relation found by \citet{2006ApJ...637..669L} (their eq. 9):
\be
\log({\rm P}_{j}) = -0.22 +
0.59\log\left(\frac{M_{bh}}{M_{\sun}}\right) + 40.48
\label{eq:4}
\ee
where ${\rm P}_{j}$ is the jet's power in CGS units, and we have adopted the value $\lambda = {\rm L}_{bol}/{\rm L}_{edd}
= 0.1$ in their equation.\\
We then embed an isothermal ISM within this NFW halo, following the
prescription given in Appendix C of \citet{2006ApJ...647..910H}: this halo is specified by the values of the central ISM density $\rho_{c0}$ and by the virial distance $r_{v}$.
\begin{figure}
%\centering
\includegraphics[scale=0.4,angle=0]{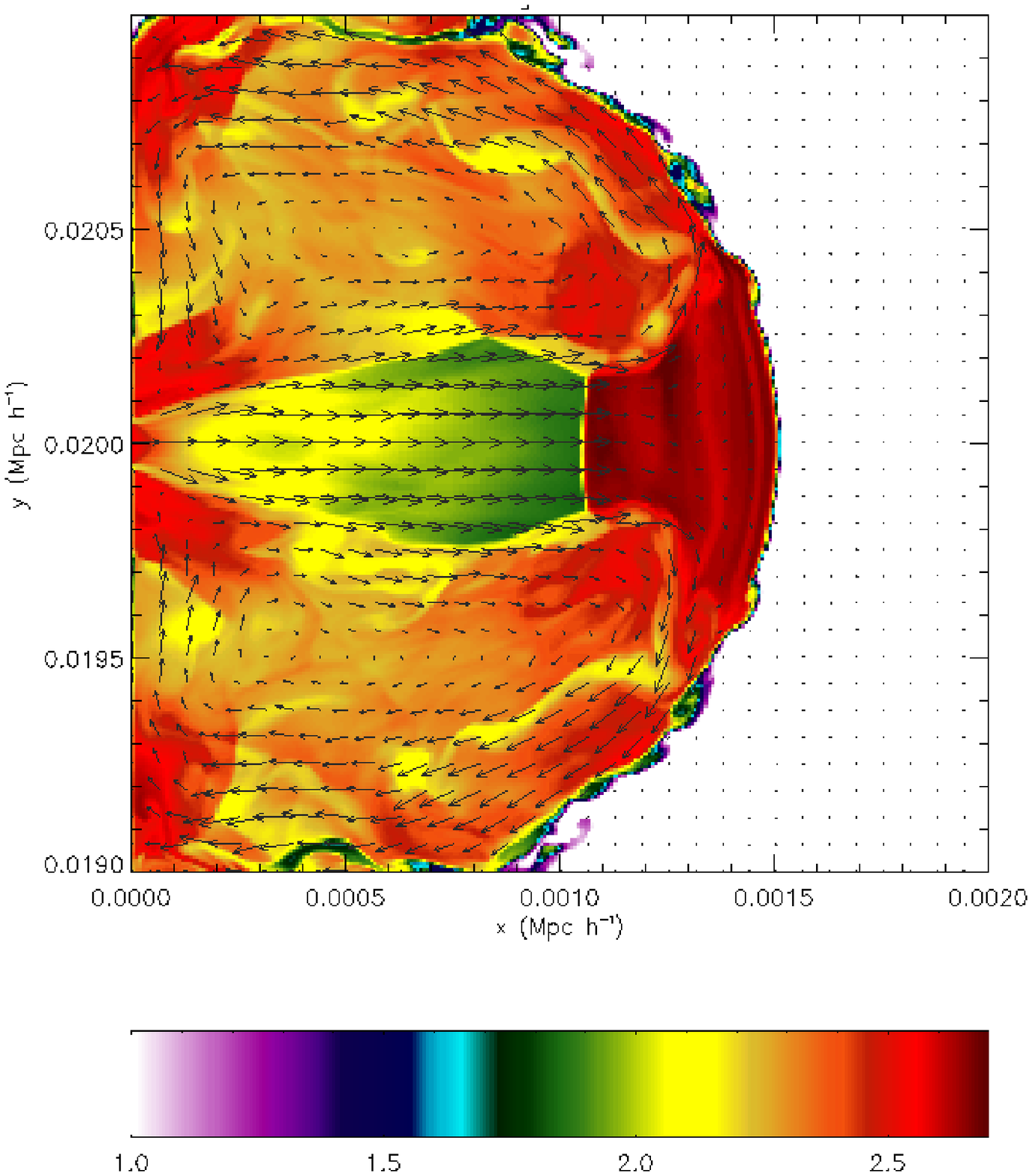}
\includegraphics[scale=0.4,angle=0]{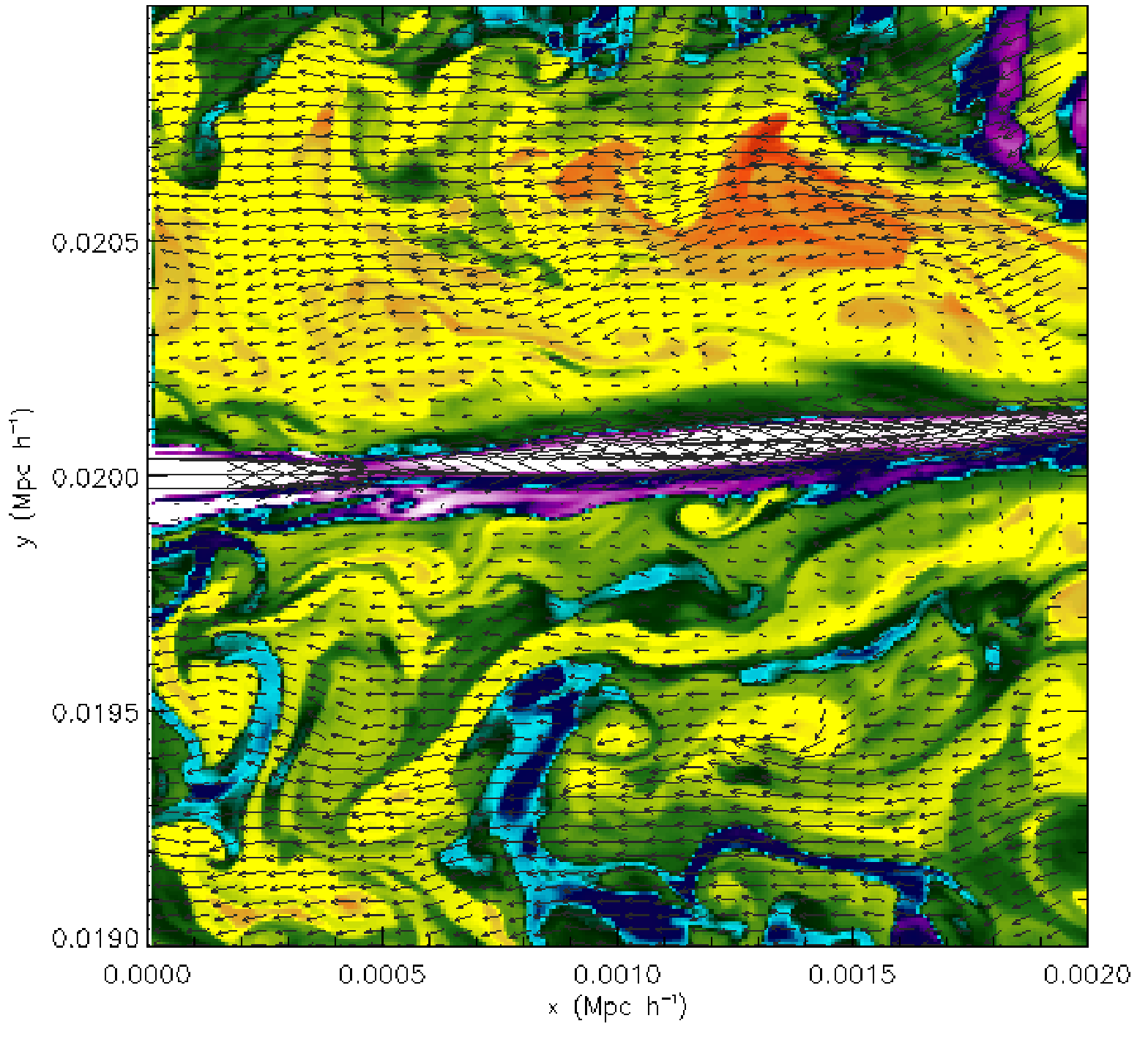}

\caption{Evolution of the velocity field for run \emph{s100av}. We
  show the significant transition of the flow near the central region,
  with a radially compressive flow at ${\rm t}\approx 13.1\times
  10^{6}$ yrs. (top) to a more laminar, longitudinal flow at ${\rm t}\approx 33.9\times
  10^{6}$ yrs. (bottom). Flow lines are superimposed on entropy
  contours. Large deviations from laminar, null vorticity flows are
  produced near discontinuities in entropy, as near the downstream
  jet's shock in the bottom figure.}
%\label{fig_fr1}
\label{fig1b}
\end{figure}
\begin{figure}
%\centering
\includegraphics[scale=0.4,angle=0]{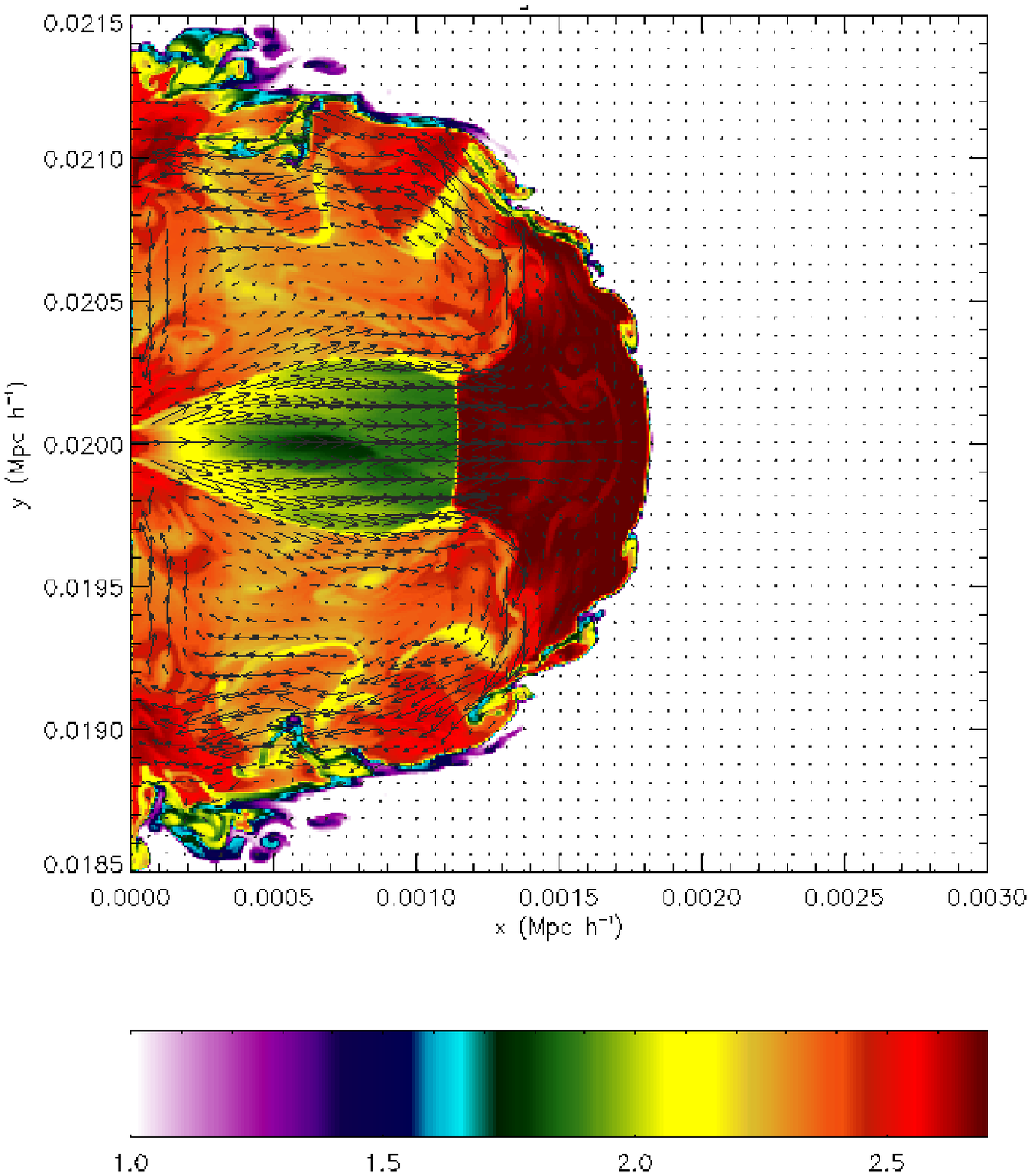}
\includegraphics[scale=0.4,angle=0]{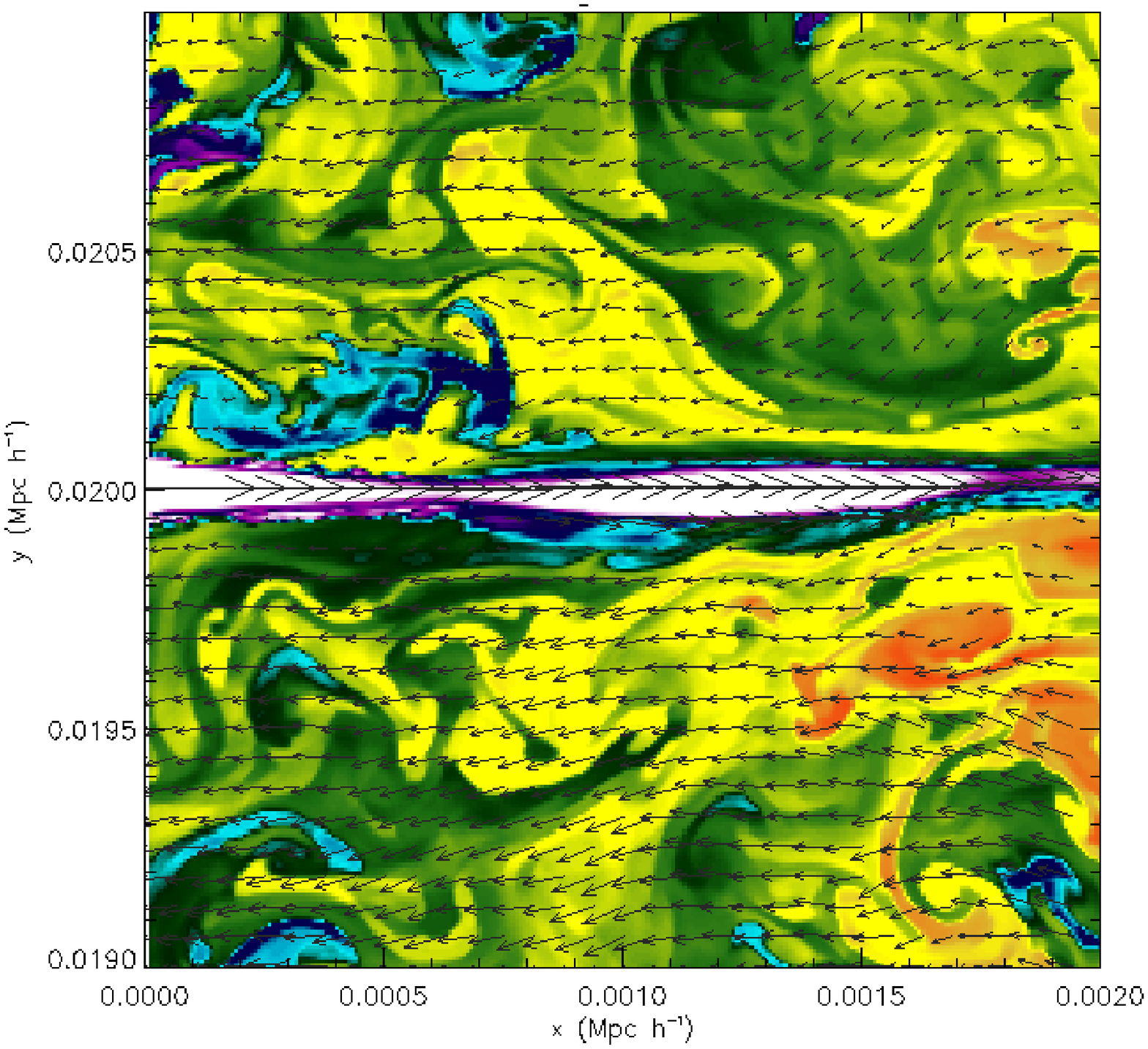}

\caption{Same as Fig. 1 for run \emph{s200av}. Top plot: ${\rm
    t}\approx 13.1\times 10^{6}$ yrs., bottom: ${\rm
    t}\approx 44.3\times 10^{6}$ yrs. }

\label{fig2}
\end{figure}
We choose the total mass of the hot, diffuse component to be equal to the
average ratio $\Omega_{b}/\Omega_{dm}$ appropriate to our chosen cosmological model, so that: $\rho_{b} = 0.174\rho_{dm}$. From this we determine the
central density $\rho_{c0}$ and scale radius of the diffuse ISM.\\
\noindent
We have performed three reference simulations, 
starting with the central values: $\sigma_{v} = 100, 200$ and $300$ km
${\rm sec}^{-1}$, and with all the other parameters determined according to
the scaling relations just quoted. In addition, for each of these
simulations, we have performed two more simulations where we have varied
\mvir (and consequently \rhoism and \rnfw) within the $\pm 1\sigma$
bounds, in order to explore a slightly larger
parameter space. The central ISM density is one of the parameters
which mostly affect the temporal evolution of the jet, as recently
shown in the simulations by \citet{2007ApJS..173...37S}.\\
\noindent 
The main input parameters of the simulations are summarized in
Table~\ref{table:1}. We assume that the jet density is a fixed fraction of the central ISM density ($\rho_{j}/\rho_{ism} = 10^{-2}$). In Table~\ref{table:2} we present some quantities related to the jet, deduced from the input parameters. It is interesting to observe that \rhoism does
not scale monotonicaly with \mvir, because it also depends on the
virial radius \rnfw. \\
\begin{table*}
 \centering
 \begin{minipage}{100mm}
  \caption{Input parameters of the simulation runs. From left to
    right, columns are as follows: run's label, halo velocity dispersion, halo
    total virial mass, central ISM density, BH mass, jet's input kinetic power.}

  \begin{tabular}{lccccc}
  \hline
   run & $\sigma_{\rm v}$ & {\rm M}$_{v}$ & $\rho_{\rm c}$ &
   {\rm M}$_{bh}$ & {\rm P}$_{k}$ \\
& ({\rm km sec$^{-1}$})&  ({\rm M}$_{\sun}$) & (cm$^{-3}$) & ({\rm M}$_{\sun}$)  & (ergs$\cdot$ sec$^{-1}$) \\
 \hline
 s100av & 100 & $3.23\cdot 10^{11}$ & $2.20$ & $8.92\times 10^{6}$ &
   $2.29\times 10^{44}$ \\
 s100p1 & '' & $6.44\cdot 10^{11}$ & $7.74$ & '' &
   '' \\
 s100m1 & '' & $1.62\cdot 10^{11}$ & $0.746$ & '' &
   '' \\
 s200av & 200 & $2.57\cdot 10^{12}$ & $2.68$ & $1.2\times 10^{8}$ &
   $1.06\times 10^{45}$ \\
 s200p1 & '' & $5.68\cdot 10^{12}$ & $13.26$ & '' &
   '' \\
 s200m1 & '' & $1.16\cdot 10^{12}$ & $0.893$ & '' &
   '' \\
 s300av & 300 & $8.62\cdot 10^{12}$ & $3.04$ & $5.49\times 10^{8}$ &
   $2.61\times 10^{45}$ \\
 s300p1 & '' & $2.03\cdot 10^{13}$ & $11.69$ & '' &
   '' \\
 s300m1 & '' & $3.66\cdot 10^{12}$ & $0.98$ & '' &
   '' \\
\hline
\end{tabular}
\end{minipage}
\label{table:1}
\end{table*}

\begin{table*}
 \centering
 \begin{minipage}{100mm}
  \caption{Main parameters of the jet, deduced from the input values. From left to
    right, columns are as follows: run's label, (spatial) average input velocity, Lorentz factor, relativistic velocity, (non-relativistic) Mach number, enthalpy flux,  and ratio between enthalpy and kinetic flux. Note that for an ideal gas the enthalpy is given by: $h = \gamma p/(\gamma-1)$, thus the enthalpy provides also a measure of jet's pressure.}
 \centering

  \begin{tabular}{lcccccc}
  \hline
   run & $\langle {\rm v}\rangle_{j}$ & $\Gamma_{j}$ &
   $\beta_{j}$ & {\rm M}$_{nr}$ & {\rm h}$_{j}$ & $P_{h}/P_{k}$ \\
& (km$\cdot$sec$^{-1}$) & - & - & - & ($10^{-16}$ ergs$\cdot$ sec$^{-1}$) & - \\
 \hline
 s100av & 217.4 & $6.74$ & $0.989$ & $11.55$ &
   $22.83$ & 0.029 \\
 s100p1 & 143.0 & $3.65$ & $0.961$ & $6.08$ &
   $80.1$ & 0.108\\
 s100m1 & 312.8 & $11.55$ & $0.996$ & $19.93$ & 
   $7.72$ & 0.010 \\
 s200av & 340.7 & $13.11$ & $0.997$ & $22.65$ &
   $27.74$ & 0.008 \\
 s200p1 & 199.3 & $5.93$ & $0.985$ & $10.12$ &
   $137.3$ & 0.039 \\
 s200m1 & 491.0 & $22.68$ & $0.999$ & $39.25$ & 
   $9.25$ & 0.003 \\
 s300av & 439.2 & $19.27$ & $0.998$ & $33.33$ &
   $31.44$ & 0.004 \\
 s300p1 & 281.0 & $9.84$ & $0.994$ & $16.95$ &
   $121.05$ & 0.014 \\
 s300m1 & 64.1 & $33.9$ & $0.999$ & $58.72$ &
   $10.14$ & 0.001 \\
\hline
\end{tabular}
\end{minipage}
\label{table:2}
\end{table*}

\noindent
As in paper I and in \citet[][hereafter paper II]{2009MNRAS.396...61T}, all the simulations were performed in a 2D box of
size 40 $h^{-1}\,$ kpc, with free flow boundary conditions. We have used
the same setup for refinement and initial mesh distribution as in the above cited papers, so we
have the same maximum resolution: $78.125\, h^{-1}$ pc. The jet is
injected with a velocity $v_{j}\equiv v_{j}(y)$ from the left
boundary, with a velocity profile given by:
\be
v_{j} = \frac{V_{j}}{\cosh\left\{\left(y-y_{j}\right)^{\alpha_{j}}\right\}}
\label{eq:5}
\ee
where we have defined a dimensionless distance $y=r/d_{j}$, $y_{j} = r_{c}/d_{j}\, r_{c}\,$, being the coordinate of the symmetry center of the profile, as we previously did \citep{2009MNRAS.396...61T}.  As this velocity profile is not truncated, we estimate the size of the jet to coincide with the typical scalelength $d_{j}$. We choose $d_j=10\, h^{-1}\, \rm{pc}$, and $\alpha_{j}=10$, as previously done by \citet{1994A&A...283..655B}. The peak velocity $V_{j}$ is determined by the total kinetic jet's power:
\be
P_{k} = \pi \rho_{j}d_{j}^{2}V_{j}^{3}\int_{-\infty}^{+\infty}dy \frac{y}{\left[\cosh\left\{\left(y-y_{j}\right)^{\alpha_{j}}\right\}\right]^{3}} \label{eq:5z}
\ee
We then see that the \emph{circulation} $\bomega\equiv\nabla\times\mathbf{v}$ for this initial velocity profile is non-zero.\\
\noindent
The jet is assumed to be non relativistic at injection, as the equations we solve are also non relativistic. As it has been previously showed by \citet{1994A&A...281L...9M, 1995ApJ...448L.105M} and by \citet{1996ASPC..100..173K}, it is not possible to find an exact relativistic anologue of a classical jet, but one can find configurations which are physically comparable. We can derive the main parameters of a relativistic jet which has the same kinetic power and radius as our jet, using eq. (21) from \citet{2007ApJS..173...37S}:
\be
P_{k,r} = 3.9\times 10^{40}\frac{\gamma}{\gamma-1}\xi\pi_{7}d_{10}^{2}\Gamma_{j}^{2}\beta_{j}\left(1+\frac{\Gamma_{j}-1}{\Gamma_{j}}\chi\right) \label{eq:5a}
\ee 
where $\gamma$ is the adiabatic index of the ISM, $\xi=p_{j}/p_{ism}$, $\pi_{7}=(p_{ism}/k)/10^{7}$, $d_{10}=(2d_{j})/(10 pc)$, $\chi = \rho_{j}c^{2}/4p_{j}$, and $\Gamma_{j} = 1/\sqrt{1-\beta_{j}^{2}}$ is the jet's Lorentz factor. Given as input $P_{k,r}$ and $d_{10}=2$, we compute the Lorentz factor (see Table~\ref{table:2}). We also compute the (non relativistic) Mach number, using eqs. (17) and (18) from \citet{2007ApJS..173...37S}:
\be
M_{nr} = \frac{2}{\gamma-1}\left(1+\frac{\Gamma_{j}-1}{\Gamma_{j}}\chi\right)\left(\Gamma_{j}^{2}-1\right)
\label{eq:5b}
\ee
We note from Table~\ref{table:2} that the jet is hypersonic at injection, and that the mass flux is very small. As we will see, the mass flow from the backflow is significantly larger: most of the flux contributing to the latter comes from ISM gas dragged by the jet itself.\\
\noindent
The total energy flux injected by the jet is given by: $P_{j} = P_{k} + P_{h}$, where $P_{k}$ was given above (eq.~\ref{eq:5z}), and the enthalpy flux is given by:
\be
P_{h} = \pi d_{j}^{2}h\langle v_{j}\rangle \label{eq:5c}
\ee
where $h$ is the specific enthalpy, which in a fully ionised jet is given by:
\be
h = \frac{\gamma}{\gamma - 1}p\label{eq:5d}
\ee
We see from the last column of Table 2 that the enthalpy flux is comparable and even much larger than the kinetic flux. Thus the entrained ISM gas gets a significant thermal energy input from the jet, in addition to the kinetic drag.

\subsection{Simulation results}
The simulations confirm the general picture of our model. Two different circulation regimes describe
 the flow within the cocoon. As initially demonstrated by
\citet{1991MNRAS.250..581F} and by \citet{2006MNRAS.368.1404A}, the
jet forms a recollimation shock along its path, when the average pressure $p_{c}$ inside the cocoon is comparable to the jet's lateral pressure. The jet then is confined towards the axis. Its injection flow is 
supersonic, and at its terminal point, near the hotspot, the jet 
moves subsonically: thus, a shock develops. However, the simulations show that the position and extent of this shock changes with time. As the character of the backflow is critically dependent on the shape and position of this recollimation shock, we will analyze here in some detail this issue.
As we can see from the upper parts of
Figs.~\ref{fig1b} and~\ref{fig2}, initially this shock is perpendicular
to the flow and has a transversal extent comparable to the cocoon's extent: also, the post-shocked gas flows along the axis, thus the gas does not gain any
circulation. However, the gas flowing off-axis eventually
encounters the density discontinuity near the HS, and starts to gain 
rotational motions. This then generates a backflow, which
transports gas in the direction opposite to that of the inflowing
jet. This backflow then changes 
direction again when it reaches the meridional plane, near the vertical axis
of the configuration. Note that all the four boundaries here are not
reflecting: the fluid can freely leave the box, thus the change of
flow direction \emph{is not} a result of particular  boundary conditions. The
resulting flow could then result in \emph{compression} of a
gaseous disk perpendicular to the jet and lying in the meridional plane.\\
\noindent
\begin{figure*}
\centering
\includegraphics[scale=0.45,angle=0]{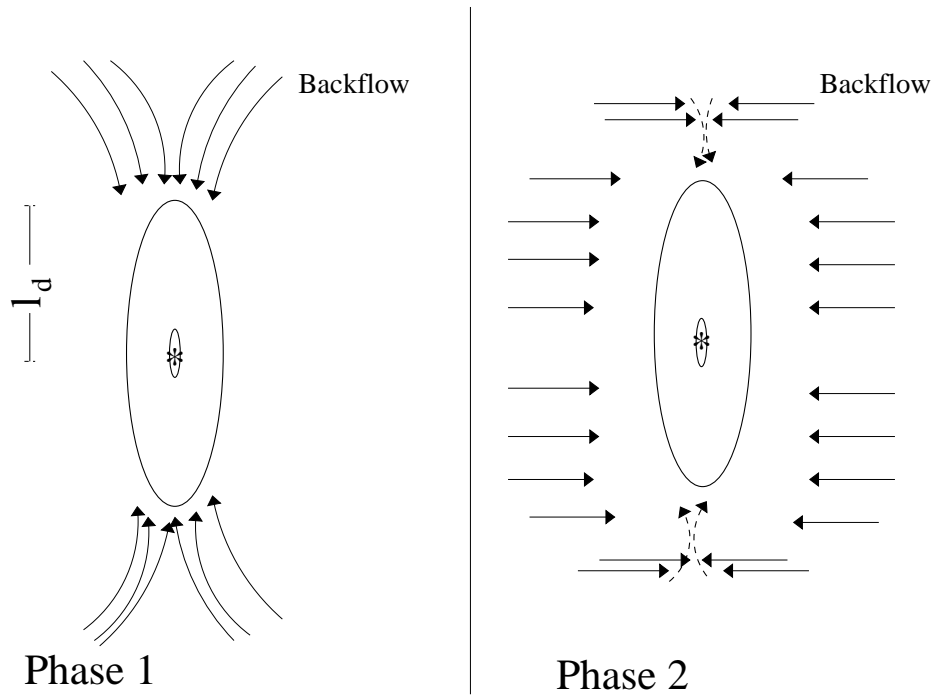}
\caption{Schematic view of the backflows during the two phases. We show a model of a realistic situation, when a two-sided jet/cocoon develops, and there are backflows from opposite directions. We assume that the circumnuclear disc has a finite radial extent, charcaterized by a finite radial scalelength $l_{d}$.}
%\label{fig_fr1}
\label{fig:bckflw:phases}
\end{figure*}
The \emph{reconfinement shock} is however unstable:
as soon as the cocoon expands, transonic turbulence develops which
destroys the shock.  The jet is now confined by the turbulent pressure
of the cocoon, but the extent of the recollimation shock is much smaller, because the cocoon's pressure $p_{c}$ steadily increases, as the jet continues to input thermal energy within the cocoon, whose expansion is slowed down by the ISM pressure. Now the pattern of the flow changes
drastically: as we can see from the bottom parts of Figs.~\ref{fig1b}
and~\ref{fig2}, the flow is mostly directed horizontally towards the accretion disc. The recollimation shock has now a much smaller transverse extension, and is located very near to the injection point, as originally shown by \citet{1991MNRAS.250..581F} and \citet{1997MNRAS.286..215K}. This general evolutionary pattern of the flow is reproduced in all the
simulations we have performed, although with some variations in the temporal
evolution and some large transient fluctuations (see Fig.~\ref{fig4}).\\
\noindent
In Figure~\ref{fig:bckflw:phases} we present a schematic summary of the resulting backflow around the circumnuclear disc, during the two main phases just mentioned. During Phase 1 the backflow mainly comes along the meridional plane, thus its angular momentum is almost zero, while during Phase 2 the backflows has a finite angular momentum component out of the plane. The part of these backflows falling outside the central nuclear disc will collide and lose most of its angular momentum, flowing then radially towards the dics, as in Phase 1. The final fate of the backflow during Phase 2, when it falls on opposite sides of the disc, will depend on whether the latter is mostly stellar or gaseous: we comment more on this in section~\ref{subs:fcbh}. During \emph{both} phases, most of the backflow will then result into a net flow of almost zero angular momentum gas towards the central regions of the AGN. In the next section we compute the actual fluxes around this circumnuclear region.

\subsection{Inflow around circumnuclear regions} \label{sec:inflow}
In so much as there exists a shock along the jet, a fraction of the backflow towards the z=0 plane is present. This is true both during the initial phase, when the shock is rather extended, and also later, when the flow is reflected by the terminal shock. This backflow mostly entails gas supplied by the lateral, subsonic jet flow. The general circulation induced by the backflow contributes a
systematic flow of low angular momentum gas in a plane orthogonal to
the jet, near the z=0 plane. It is then interesting to compute the mass flux contributed by this backflow to the very central region, near the central Black Hole.\\ 
\noindent
If an accretion disk is present around the BH, it would lie in a plane
perpendicular to the jet, and its structure will probably be affected
by the backflow. In Fig.~\ref{fig5} we show the average mass flow
within an annular region having an internal radius $r_{in}=1.5\,$ pc, a radius $r_{c} = 20\,$ pc and height $h_{c}=2\,$ pc, for the three main runs. The mass flux across the lateral surface of the disc will be given by:
\be
\dot{M}_{r} = 2\pi r_{c}\int_{0}^{h_{c}}dz \rho_{f}v_{r} 
\label{eq:simn1:1}
\ee
assuming axial symmetry.  There will also be a flux across the upper surface of the disc:
\be
\dot{M}_{z} = 2\pi\int_{r_{in}}^{r_{c}+d_{c}}dr r\rho_{f}v_{z} 
\label{eq:simn1:2}
\ee
We show in Figure~\ref{fig6} the evolution of the total mass flux $\dot{M}_{r} +\dot{M}_{z}$. 
It is interesting to observe that the accretion rates contributed by the backflow
all seem to vary on similar time scales, of the order of $3.5-4\times
10^{7}\,$ yrs., during which the accretion rate can reach peak values in
excess of 1 $\rmn{M}_{\sun}\,\rmn{yr}^{-1}$, i.e. the typical
accretion rates required to support activity in QSO and Seyferts \citep[see e.g.][Table 1]{2006LNP...693..143J}. This time-scale seems to be almost independent of the ratio between the bulge and BH masses, but we see
large variations induced by variations in the local ISM density (Fig.~\ref{fig6}): systems
where the local ISM density is larger seem also to
have  larger accretion rates, with peak values exceeding 4-5
$\rmn{M}_{\sun}\,\rmn{yr}^{-1}$. This suggests that the accretion
rate can be very sensitive to the occurrence of recent episodes of wet
mergers, during which the gas density can easily vary by $\pm
1\sigma\,$ w.r.t. the average values.\\
\begin{figure}
%centering
\includegraphics[scale=0.4,angle=0]{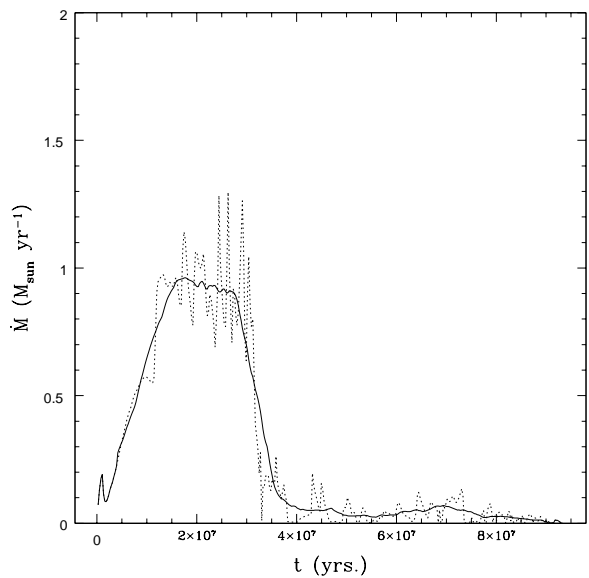}
\caption{Mass flow in the circumnuclear region for run
  \emph{sim200av}. We show the averaged flux (continuous line) and the
  instantaneous flux (dotted line), both computed by adding the contributions from eqs.~\ref{eq:simn1:1}-~\ref{eq:simn1:2}, as described in section~\ref{sec:inflow}.. The transient fluctuations are caused
  by the intermittent behaviour of the flux inside the cocoon, during
  the evolved phase of the flow.}
%\label{fig_fr1}
\label{fig4}
\end{figure}
\noindent
It is also interesting to observe that the accretion is highly
intermittent, as can be appreciated by looking at
Fig.~\ref{fig4}. The beginning of the intermittency coincides with
the disappearance of the reconfinement shock, and marks the onset of 
fully developed, transonic turbulence within the cocoon. If part or all of this backflow feeds the central BH, the intermittency could then translate into an intermittency of the feeding mechanism, and result in a random orientation of the jet axis with the structure of the  surrounding galaxy \citep{2000ApJ...537..152K}.
\\
\noindent
Finally, we have also plotted in Fig.~\ref{fig6} the predictions about the mass flow rates from the model that we developed in the previous sections. The model seems to adequately describe  the results from the simulations. The \emph{quantitative} agreement is quite good, at least during the phase dominated by the regular backflow, i.e. the only phase which we have modelled. We have adopted the temporal
dependence using the expansion law given by
\citet{1997MNRAS.286..215K}, for the input values of $P_{k}$ and
$n_{ism}$ as given in Table 1 (we consider only the fiducial
models). We notice that the
largest deviations are for smaller values of $\sigma_{v}$, when the expansion is slower and the deviations from the self-similar model by Kaiser \& Alexander are larger.\\
\begin{figure}
\centering

\includegraphics[scale=0.4,angle=0]{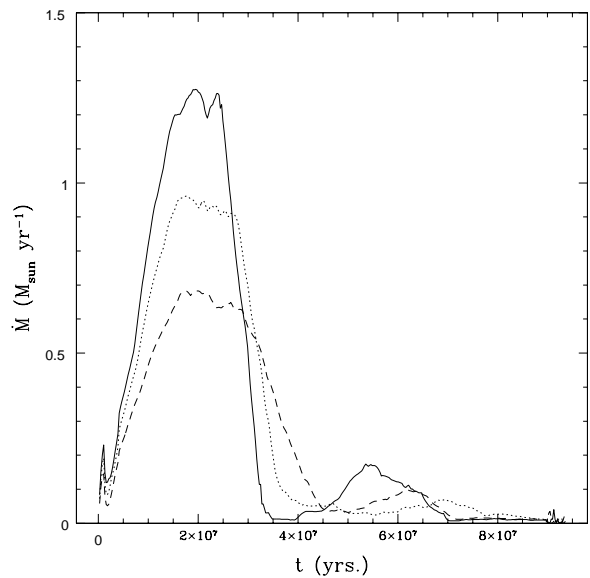}
\caption{The average mass flow rate around a circumnuclear region,
  with a diameter $d = 20$ pc centered around the circumnuclear
  disk. The flow is computed as described in Figure~\ref{fig4}.} We show only the results for runs \emph{s300av} (continuous
  line), \emph{s200av} (dotted line), and \emph{s300av}
  (dashed). Time scales are almost constant, while the maximum of the
  curve seems to depend on $P_{k}$. 
\label{fig5}
\end{figure}
 \begin{figure}
\centering
\includegraphics[scale=0.4,angle=0]{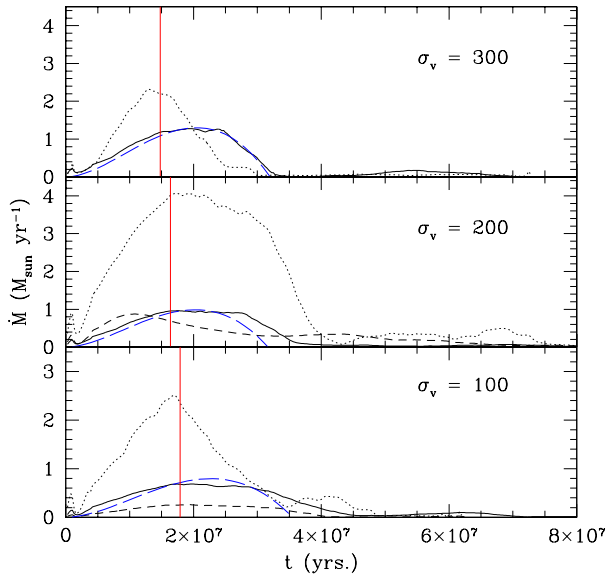}
\caption{Mass inflow rates around a central disk region, computed as described in Figure~\ref{fig4}. The continous
   lines are for simulations \emph{s(100,200,300)av}, dotted lines for
   \emph{s(100,200,300)p1}, dashed lines for
   \emph{s(100,200,300)m1}. The blue dashed lines are the fits from
   the model given in the text. The red vertical lines mark the epoch
   when the reconfinement shock is destroyed and the jet starts to
   propagate freely, being only confined by the cocoon's ram pressure.}
%\label{fig_fr1}
\label{fig6}
\end{figure}
\noindent
\begin{table*}
 \centering
 \begin{minipage}{100mm}
  \caption{Mass inflow rates. We show the input mass flow rate, $\langle \dot{\rm M}\rangle_{in}$ (column 2), and the (averaged in time) flow entrained within the backflow around the circumnuclear region (column 3). Flows are in units of ${\rm M}_{\sun} {\rm yr.}^{-1}$.}
 \centering

  \begin{tabular}{lcc}
  \hline
   run & $\langle \dot{\rm M}\rangle_{in}$ & $\dot{\rm M}_{cn}$ \\
& &  \\
 \hline
s100av & $2.67\cdot 10^{-4}$ & $3.12\cdot 10^{-1}$ \\
s100p1 & $6.7\cdot 10^{-4}$ & $4.61\cdot 10^{-1}$ \\
s100m1 & $1.41\cdot 10^{-4}$ & -  \\
s200av & $5.52\cdot 10^{-4}$ & $2.73\cdot 10^{-1}$  \\
s200p1 & $1.6\cdot 10^{-3}$ & $1.33\cdot 10^{-1}$  \\
s200m1 & $2.65\cdot 10^{-4}$ & $2.71\cdot 10^{-1}$  \\
s300av & $8.06\cdot 10^{-4}$ & $2.26\cdot 10^{-1}$  \\
s300p1 & $1.98\cdot 10^{-3}$ & $7.65\cdot 10^{-1}$  \\
s300m1 & $3.79\cdot 10^{-5}$ & $1.22\cdot 10^{-1}$  \\
\hline
\end{tabular}
\end{minipage}
\label{table:3}
\end{table*}
In Table 3 we compare the input mass flow with the flow entrained by the jet: the latter is about two orders of magnitude larger than the first. The combined action of kinetic and enthalpy pressure exerted by the jet on the ISM is able to create a circulation which drags a significant amount of mass flow towards the central regions. Note that this however takes place on a time scale of the theorder of few times $10^{7}\,$ yrs.
\noindent
\subsubsection{Feeding the central BH} \label{subs:fcbh}
We  now briefly consider the final fate of the inflowing gas. As we have seen, this gas has  very low angular momentum, at variance with most of the gas in the outer regions of the circumnuclear disc.
\noindent
However, the final fate of this gas will primarily depend on the structure of the circumnuclear disc around the central BH, and in particular on whether this disk has a pressure which can signficantly resist the inflow. Since the seminal paper by \citet{1989ApJ...341..685S}, it has been  suggested that on scales larger than 10-50 pc,  the circumnuclear discs around AGNs are prone to be unstable to gravitational fragmentation \citep{1994mtia.conf...23B, 2003MNRAS.339..937G, 2002ApJ...566L..21W, 2004cbhg.symp..186W, 2005MNRAS.362..983T, 2007ApJ...671.1264E, 2008ApJ...685L..31E}. Recently, it has been suggested that star formation within this disc can provide a  self-regulatory mechanism which has a two-fold action: heating the disc, thus halting star formation \citep{1999A&A...344..433C, 1999Ap&SS.265..501C}, and feeding the central BH through massive stellar winds \citep{2005ApJ...630..167T, 2009arXiv0907.1002N}. More recent  observations with resolution on parsec scales are now confirming the presence of dense star clusters in the central parts of bulges which are also hosting AGNs \citep{2007ApJ...671.1388D, 2008ApJ...678..116S, 2008AJ....135..747G, 2009arXiv0907.5250G}, thus providing some evidence for the existence of  gas and especially stars  within these circumnuclear discs, from a few to a few hundreds of parsecs.\\
\noindent
Thus, we expect that the inflowing gas provided by the large-scale backflow will mostly ``see'' a stellar disc, and will interact with the stars by creating local Bondi-like accretion flows around them. The cross sections for Bondi flows are very small (of the order of few stellar radii), and we can then reasonably predict that most of the inflow will be able to reach the inner gaseous, high-density ($\rm{n}_{e} > 10^{8}\, \rm{cm}^{-3}$) disc, thus providing  significant feeding of the latter. In a forthcoming paper (Antonuccio-Delogu \& Silk 2010, in preparation), we will analyse this interaction in detail to understand quantitatively how much the inflow can contribute to  actually feeding  the AGN. 

\section{Conclusions}
The development of a backflow within the cocoon  that results from the interaction of a relativistic AGN jet with the surrounding ISM has been previously studied by \citet{1999MNRAS.305..707K}. Our main focus in this work has been to show that a significant contribution to the development of circulation within the cocoon comes from the presence of naturally developing density and entropy gradients.\\
\noindent
 Our calculation can have direct implications for the structure of AGNs' and of their circumnuclear discs. One of the best studied cases is the ionised circumnuclear disc in the central region of M87 \citep{1994ApJ...435L..27F}. HST spectroscopic observations have shown show that this disc is shock-excited \citep{1997ApJ...490..202D}, and its emission lines are probably originating from accretion of gas external to the disc (see however the critical remarks by \citet{2003ApJ...584..164S}). It is then possible that the backflow we have studied in this work could contribute to the accretion shock. We plan to present a detailed modelling of the contribution of this flow to the pbserved spectrum in a future work.\\
\noindent
One of the first questions concerns the stability of this flow, and a useful hint towards an answer comes from the simulations we have performed. Although a  linear stability analysis would also be possible, we see that the flow is not heavily affected by fluctuations induced by turbulence within the cocoon. However, when the recollimation shock is destroyed by instabilities within the cocoon,  the main source of circulation, which arises from discontinuities within the HS, is destroyed, and the backflow changes abruptly, becoming almost totally longitudinal. The shape of the jet now changes: the recollimation shock forms again, but much nearer the injection point, as originally predicted by \citet{1991MNRAS.250..581F} and \citet{1997MNRAS.286..215K}, and a new shock forms immediately before the hot spot, from which now a backflow develops which flows almost parallel to the confined jet. Also this backflow contributes very low angular momentum gas to the central regions of the AGN, with average mass inflow rates of the order of a few $\rm M_{\odot}\, {yr.}^{-1}$, on time-scales of the order of a few times $10^{7}$ yrs, as  required in order to feed an AGN.\\
\noindent
In summary, we can distingusih two main phases in the development of flow within the cocoon:
\begin{itemize}
\item \emph{Phase 1}: A large scale backflow develops, fed by some gas originating in the lateral regions of the Hot Spot, and flowing along the inner region of the contact discontinuity, up to the meridional plane of the cocoon, and eventually all the way down towards the AGN;
\item \emph{Phase 2}: When the jet becomes pressure confined, gas originating from the HS flows back longitudinally towards the AGN.
\end{itemize}
During both phases the backflow provides (almost) zero angular momentum gas to the central region of the AGN.
\\
\noindent
We have seen that the backflow originates when the subsonic flow of the outer parts of the jet hits some shock, and then gains angular momentum. Thus, this flow is not arising from some turbulent entrainment, as is seen when part of the 3D turbulent energy is dissipated through shear and engenders a regular motion of entrained gas. This is the reason why this backflow is also manifest in 2D simulations, as those we have studied in this work.\\
\noindent
Competing mass inflow mechanisms,  in particular  low angular momentum gas driven inwards by a nuclear spiral disk,  can power nuclear starbursts \citep{2008ApJ...675L..17F} but 
give substantially smaller mass inflow rates within the central few parsecs \citep[e.g.][]{2009ApJ...702..114D}.
\\
\noindent
It is interesting to observe that the inflow rates appear to depend strongly on the local ISM density: more massive and denser galaxies host stronger backflows, because there is more matter which is dragged in by the jet. The average mass inflow rates in the circumnuclear regions are always of the the order of $10^{-1}\, {\rm M}_{sun}\, {\rm yr.}^{-1}$, although they do not scale only with the jet's input power, but seem to be strongly affected by the ISM density, as can be seen from Table 3. The kinetic pressure of more powerful jets is typically much larger than the ram pressure from the gas: only within the turbulent region of the cocoon, where the backflow is absent, does one reach  approximate equipartition (paper I). \\
This backflow could thus provide the main source for  feeding the central AGN with very low angular momentum gas, which would eventually reach the BH and feed the jet. We see that, when this backflow is present,  the jet power increases with time, and the cocoon expands more rapidly. Ultimately, this would have the effect of accelerating the expansion of the cocoon and destroying the backflow, thus halting the feeding of the AGN. From Figure 9 we see that the mass inflow rate initially increases, thus providing more fuel to the AGN, but then decreases, as the backflow now originates from gas flowing near the hotspot, whose energy decreases as the cocoon expands. One can presume that a fraction of the inflowing gas will be able to reach the AGN and feed it, possibly increasing $P_{k}$. This however depends on the fate of the inflowing gas when it meets the inner accretion disc and the stellar disc, i.e. on phenomena taking place on parsec scales, much smaller than those resolved by the simulations we have presented in this paper.\\
\noindent
We have thus described a \emph{self-regulation mechanism}, which does not directly invoke any starburst in the circumnuclear disk to feed the AGN. The role of starbursts  in AGN feeding is that of supplying low angular momentum gas from AGN winds, which can eventually reach the central black hole. In this backflow model, the low angular momentum gas is provided from the jet itself: however, more detailed modelling of its interaction with the accretion disk is needed to understand quantitatively how much this inflow can contribute to the feeding of  the AGN.

\section*{Acknowledgments}
We are very pleased to acknowledge the many useful comments from the referee, prof. G. Bicknell, which helped to improve the presentation of this work.\\
The work of V. A.-D. has been supported by the European Commission,
under the VI Framework Program for Research \& Development, Action
``{\em Transfer of Knowledge}'' contract
MTKD-CT-002995 (''{\em Cosmology and Computational Astrophysics at
  Catania Astrophysical Observatory}''). This work makes use of results produced by the PI2S2 Project managed
by the Consorzio COMETA, a project co-funded by the Italian Ministry
of University and Research (MIUR) within the {\em Piano Operativo Nazionale 
"Ricerca Scientifica, Sviluppo Tecnologico, Alta Formazione" (PON
2000-2006)}. More information is available at http://www.pi2s2.it (in
italian) and http://www.trigrid.it/pbeng/engindex.php .
%\bibliographystyle{/home/van/tex/mnras/mn2e}
%\bibliography{/data/uist/van/papers/ref_j/jet_ism/biblio}
\bibliographystyle{mn2e}
\bibliography{biblio}

\appendix

%\section[]{Oblate spheroidal coordinates} \label{append_0}

\section[]{A vorticity conservation theorem for 2D motion} \label{append_1}
In the particular case of a 2D motion, when the circulation
$\mathbf{\omega}$ is everywhere perpendicular to the velocity field,
it is possible to show that the quantity
$\mid\mathbf{\omega}\mid/\rho$ is constant. For a general 2D
compressible motion the evolution of the vorticity is given by
\citep[eq. 1.46]{1962cole..book.....P}:
\be
\frac{d\boldsymbol{\omega}}{dt} = -\boldsymbol{\omega}\nabla\cdot\mathbf{v}
\label{app1:eq1}
\ee
where, as usual, the total time derivative denotes the
\emph{convective} derivative.\\
From the equation of continuity we also have:
\[
\frac{d\rho}{dt} = -\rho\nabla\cdot\mathbf{v},
\] 
thus: $d\ln\rho/dt = -\nabla\cdot\mathbf{v}$, and, after substituting in
eq.~\ref{app1:eq1}, we eventually obtain:
\[
\frac{d\boldsymbol{\omega}}{dt} = \boldsymbol{\omega}\frac{d\ln\rho}{dt} 
\]
Dot multiplying both sides by $\boldsymbol{\omega}$, we arrive at:
\[
\frac{1}{2}\frac{d}{dt}\ln\mid\boldsymbol{\omega}\mid^{2} =
\frac{d}{dt}\ln\mid\boldsymbol{\omega}\mid = \frac{d}{dt}\ln\rho
\]
which can be rewritten as:
\be
 \frac{d}{dt}\ln\left(\frac{\mid\boldsymbol{\omega}\mid}{\rho}\right) = 0
\label{app1:eq2}
\ee
Thus, the quantity $\mid\boldsymbol{\omega}\mid/\rho$ is advected along the
streamlines.

\section[]{Solutions of flow equations} \label{append_2}
We will here describe a scheme to derive approximate solutions of the flow equations along the backflow. As described in the text, we distinguish three regions in the backflow: a) the Hot Spot, b) the bow shock, c) the meridional region. We will now shortly describe the solutions in each of these three regions.

\begin{figure}
\centering

\includegraphics[scale=0.4,angle=0]{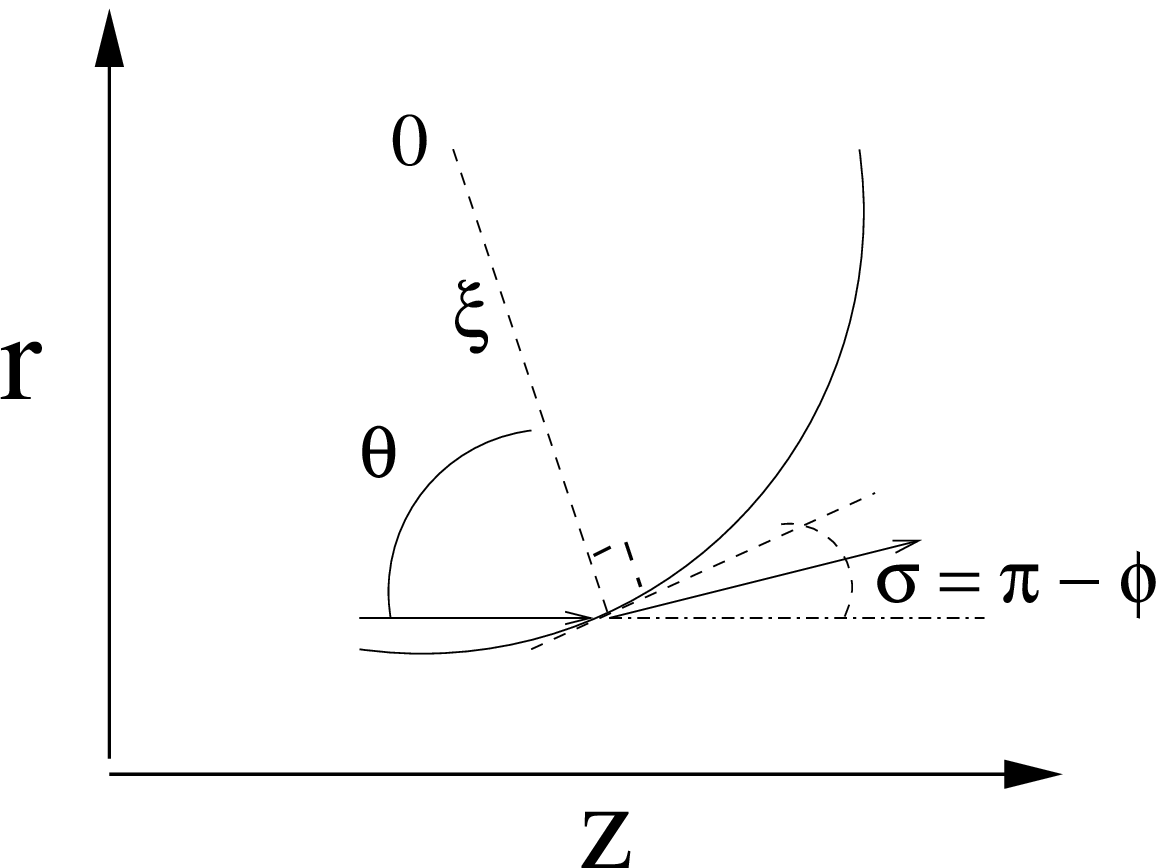}
\caption{Coordinate setup used to solve the circulation equation near the HS region.} 

\label{fig7}
\end{figure}

\subsection[]{Streamlines across the HS} \label{append_2a}
We make reference to the geometry described in Figure~\ref{fig7}. We define the polar angle $\phi$ (not shown in figure) as the angle between the r-axis and the position vector $\xi$. Introducing an auxiliary angle $\theta$, we see from the figure that $\sigma = \pi - (\pi/2+\theta) = \pi/2-\theta$, and: $\phi = \pi + (\pi/2 -\theta) = 3\pi/2 - \theta$; thus, we have: $\sigma = \phi - \pi$.  We will find useful to compute the circulation in a reference frame centered at the origin of the (assumed) spherical region bounding the sides of the HS. In a 2D spherical coordinate system, the out-of-plane component of the  circulation is then:
\be
\omega = \frac{1}{\xi}\left[\frac{\partial }{\partial\xi}\left(\xi v_{\phi\mid d}\right)-\frac{\partial v_{\xi\mid d}}{\partial\phi}\right]
\label{app2:eq1}
\ee
The continuity condition across the normal gives:
\be
\rho_{u}v_{z\mid u}\cos\theta = \rho_{d}v_{\xi\mid d}
\label{app2:eq2}
\ee
where the subscripts ``$u$'' and ``$d$'' denote quantities evaluated in the up- and downstream regions, respectively. After substituting $v_{\xi\mid d}$ from the latter equation in eq.~\ref{app2:eq1}, and assuming that the downstream region is homogeneous, we see that only the first term in the right-hand-side of eq.~\ref{app2:eq1} is different from zero.\\
We now apply Crocco's theorem, making use of the expression for the circulation after a curved shock layer, derived by \citet[][eq. 8]{1958ZAMP...9..637}:
\be
\frac{1}{\xi}\frac{\partial }{\partial\xi}\left(\xi v_{\phi\mid d}\right) = \frac{2v_{z\mid u}}{r_{s}\left(\gamma + 
  1\right)}\frac{\left(M_{n}^{2}-1\right)\mid\cos(\phi)\mid}{M_{n}^{2}[2+\left(\gamma-1\right)M_{n}^{2}]}
\label{app2:eq3}
\ee
We now integrate both terms of this equation across a layer of width $\Delta r$, starting from the layer at $\xi=r_{s}$, thus obtaining:
\[
(r_{s}+\Delta r)v_{\phi\mid d}(r_{s}+\Delta r) - r_{s}v_{\phi\mid d}(r_{s}) =
\]
\be
= \left(2r_{s}+\Delta r\right)\Delta r\frac{2v_{z\mid u}}{r_{s}\left(\gamma + 
  1\right)}\frac{\left(M_{n}^{2}-1\right)\mid\cos(\phi)\mid}{M_{n}^{2}[2+\left(\gamma-1\right)M_{n}^{2}]}
\label{app2:eq4}
\ee
%where we have dropped the subscript ``$d$'' in the left-hand side. 
Passing to the limit $\Delta r\rightarrow 0$, after having expanded $v_{\phi}(r_{s}+\Delta r)$ in Taylor series, we finally obtain:
\be
v_{\phi}(r_{s}) + r_{s}\frac{\partial v_{\phi}}{\partial\xi}\mid_{\xi=r_{s}} = \frac{4v_{z\mid u}}{\left(\gamma + 
  1\right)}\frac{\left(M_{n}^{2}-1\right)\mid\cos\phi\mid}{M_{n}^{2}[2+\left(\gamma-1\right)M_{n}^{2}]}
\label{app2:eq5}
\ee
As it is evident from this equation, we have yet to specify the gradient of the angular velocity across the boundary: however, we will now show that, by applying the continuty equation, the latter condition can be replaced by an initial condition on the circular velocity \vphi . Until this point we have only used the equation for the circulation, but we can get an additional constraint using the continuity equation:
\be
\nabla\cdot\left(\rho\mathbf{v}\right)\equiv \frac{1}{\xi}\frac{\partial}{\partial\xi}\left(\xi\rho_{d} v_{\xi\mid d}\right) + \frac{1}{\xi}\frac{\partial}{\partial\phi}\left(\rho_{d} v_{\phi\mid d}\right) = 0
\label{app2:eq6}
\ee
Inserting in the latter equation the expression for $\rho_{d}v_{\xi}$ from eq.~\ref{app2:eq2}, and using the homogeneity in the downstream HS region ($\partial\rho_{d}/\partial\phi = 0$), we eventually get the following equation for the angular velocity:
\be
\frac{1}{\xi}\cdot\rho_{u}v_{z\mid u}\sin\phi + \frac{\rho_{d}}{\xi}\frac{\partial v_{\phi}}{\partial\phi} = 0
\label{app2:eq7}
\ee
which immediately provides the following solution:
\be
v_{\phi\mid d} = \frac{\rho_{u}}{\rho_{d}}v_{z\mid u}\left(1+\cos\phi\right) + v_{\phi}\left(\xi,\phi=\pi\right)
\label{app2:eq8}
\ee
Note that $\cos\phi = -\sin\theta$, thus, imposing that eqs.~\ref{app2:eq7} and~\ref{app2:eq8} describe the same solution, we obtain:
\[
\frac{\rho_{u}}{\rho_{d}} = \frac{2}{\gamma+1}\cdot\frac{\left(M_{n}^{2}-1\right)\sin(\theta)}{M_{n}^{2}[2+\left(\gamma-1\right)M_{n}^{2}]}
\]
and:
%\[
%v_{\phi}\left(\xi,\phi=\pi\right) = -\frac{\rho_{u}}{\rho_{d}}v_{z\mid u} + v_{\phi}\left(\xi,\phi=\pi\right) + r_{s}\frac{\partial v_{\phi}}{\partial\xi}\mid_{\xi=r_{s}}
%\]
%[07-09-2009] - Credo che ci sia un'errore nell'eq. precedente
\[
v_{\phi}\left(\xi,\phi=\pi\right) = -\frac{\rho_{u}}{\rho_{d}}v_{z\mid u} + r_{s}\frac{\partial v_{\phi}}{\partial\xi}\mid_{\xi=r_{s}}
\]
\subsection[]{Solution along the bow shock.} \label{append_2b}
To describe the motion in the region of the bow shock, we will make use of a more convenient oblate spheroidal coordinate system, as defined previously in section~\ref{sect:2}.
%, as defined in Appendix~\ref{append_0}. 
We will also assume that $v_{\psi}\gg v_{\xi}$, i.e. that the expansion of the cocoon is negligible w.r.t the meridional circulation. Inserting this into the continuity equation (eq.~\ref{app0:eq3}) we can integrate the latter to obtain:
\be
h_{\xi}\rho v_{\psi} - h_{\xi}\rho v_{\psi}\mid_{\psi=\psi_{0}} = p(\xi)
\label{app3:eq1}
\ee
where $p(\xi)$ is an integration constant. Using this latter equation, we can the express the density as:
\be
\rho(\xi,\psi) = \frac{q(\xi) + p(\xi)}{h_{\xi}v_{\psi}}
\label{app3:eq2}
\ee
where we have defined: $q(\xi) = h_{\xi}\rho v_{\psi}\mid_{\psi=\psi_{0}}$.
We now make use of the law of vorticity conservation (eq.~\ref{app1:eq2}) to deduce that:
\be
\omega \approx \frac{1}{h_{\xi}h_{\psi}} \frac{\partial}{\partial \xi}\left( h_{\xi}v_{\psi}\right) = c_{0}(a)\rho 
\label{app3:eq3}
\ee
where the function $c_{0}(a)$ is determined from the initial conditions, i.e. from the typical vorticity and density at the injection point. Substituting the density from eq.~\ref{app3:eq2} in eq.~\ref{app3:eq3} we obtain:
\[
\frac{1}{h_{\xi}^{2}}\frac{\partial}{\partial \xi}\left( h_{\xi}v_{\psi}\right) = c_{0}(a)\frac{q(\xi) + p(\xi)}{h_{\xi}v_{\psi}}
\]
This can be integrated to give:
\be
h_{\xi}^{2}v_{\psi}^{2} = h_{\xi}^{2}v_{\psi}^{2}\mid_{\xi=\xi_{0}} + a^{2}\left[ u(\xi) + \left\{ v(\xi)-v(\xi_{0})\right\}\sin^{2}\psi\right] - a^{2}u(\xi_{0})
\label{app3:eq4}
\ee
where we have made usage of the expression for the scale factor $h_{\xi,\psi}$ given in section~\ref{sect:2}, and introduced the definitions:
%Appendix~\ref{append_0}
\[
r(a,\xi)\equiv c_{0}(a)\left[q(\xi)+p(\xi)\right]
\]
and:
\[
u(\xi) = 2\int d\sigma r(\sigma)\sinh^{2}\sigma, \hspace*{1cm} v(\xi) = 2\int d\sigma r(\sigma)
\]
Now, defining: $\rho_{0}\equiv\rho(\xi,\psi_{0})$, and: $v_{0}\equiv v_{\psi}(\xi,\psi_{0})$, we see that:
\[
q(\xi) = \rho_{0}v_{0}a\left(\sinh^{2}\xi+\sin^{2}\psi_{0}\right)^{1/2}
\]
Without loss of generality, we now assume: $p(\xi)\equiv 0$. Thus: $r(\xi) = c_{0}(a)q(\xi)$, and, subsituting in the expressions above, we have:
\be
u(\xi) = \frac{1}{6}c_{0}\rho_{0}v_{0}a\left[\cosh(3\xi) - 9\cosh(\xi)\right]_{\xi_{i}}^{\xi}
\label{app3:eq5}
\ee
and:
\be
v(\xi) = 2c_{0}\rho_{0}v_{0}a\cosh(\xi)\mid]_{\xi_{i}}^{\xi}
\label{app3:eq6}
\ee
In deriving these expressions, we have assumed that $\sin^{2}\psi_{0}\sim 0$, i.e. that the injection point of the bow shock is very near to the equatorial plane, a reasonable assumption if the Hot Spot lies near the tip of the expansion zone. Inserting eqs.~\ref{app3:eq5} and~\ref{app3:eq6} into the general expression eq.~\ref{app3:eq4} we eventually get the solution:
\[
v_{\psi}^{2} = \frac{ac_{0}\rho_{0}v_{0}}{\sinh^{2}\xi + \sin^{2}\psi}\cdot
\]
\be
\left[\frac{1}{6}\cosh(3\xi)-\frac{3}{2}\cosh\xi+2\cosh\xi\cdot\sin^{2}\psi\right]_{\xi_{i}}^{\xi}
\label{app3:eq7}
\ee
where, without loss of generality, we have assumed that: $h_{\xi}^{2}v_{\psi}^{2}\mid_{\xi_{i}} = 0$.
\label{lastpage}

\end{document}